\documentclass[conference]{IEEEtran}
\IEEEoverridecommandlockouts
\usepackage{cite}
\usepackage{amsmath,amssymb,amsfonts}
\usepackage{algorithmic}
\usepackage{graphicx}
\usepackage{textcomp}
\usepackage{xcolor}
\usepackage{booktabs}
\usepackage{colortbl}
\usepackage{multirow}
\usepackage{bm}
\usepackage[justification=centering]{caption}
\usepackage{balance}
\usepackage{float}
\usepackage{comment}
\usepackage{makecell}
\usepackage{hyperref}
\def\BibTeX{{\rm B\kern-.05em{\sc i\kern-.025em b}\kern-.08em
    T\kern-.1667em\lower.7ex\hbox{E}\kern-.125emX}}
\begin{document}

\title{Automatic Identification and Extraction of Assumptions on GitHub}


\author{

\IEEEauthorblockN{Chen Yang\IEEEauthorrefmark{2}\IEEEauthorrefmark{3}, Zinan Ma\IEEEauthorrefmark{2}, Peng Liang\IEEEauthorrefmark{4}\IEEEauthorrefmark{1}, Xiaohua Liu\IEEEauthorrefmark{2}\thanks{DOI reference number: 10.18293/DMSVIVA2023-071}}
    \IEEEauthorblockA{
    \IEEEauthorrefmark{2}School of Artificial Intelligence, Shenzhen Polytechnic, Shenzhen, China\\
    \IEEEauthorrefmark{3}State Key Laboratory for Novel Software Technology, Nanjing University, Nanjing, China\\
    \IEEEauthorrefmark{4}School of Computer Science, Wuhan University, Wuhan, China\\
    \{yangchen, 21680260, lxh\}szpt.edu.cn, liangp@whu.edu.cn
    }
}

\maketitle

\begin{abstract}
In software development, due to the lack of knowledge or information, time pressure, complex context, and many other factors, various uncertainties emerge during the development process, leading to assumptions scattered in projects. Being unaware of certain assumptions can result in critical problems (e.g., system vulnerability and failures). The prerequisite of analyzing and understanding assumptions in software development is to identify and extract those assumptions with acceptable effort. 
In this paper, we proposed a tool (i.e., Assumption Miner) to automatically identify and extract assumptions on GitHub projects. To evaluate the applicability of Assumption Miner, we first presented an example of using the tool to mine assumptions from one large and popular deep learning framework project: the TensorFlow project on GitHub. We then conducted an evaluation of the tool. The results show that Assumption Miner can effectively identify and extract assumptions from the repositories on GitHub.
\end{abstract}

\begin{IEEEkeywords}
Assumption, GitHub, Mining Software Repositories
\end{IEEEkeywords}

\section{Introduction} \label{Introduction}
Assumptions in the field of software development is a broad topic: different types of assumptions (e.g., requirement assumptions \cite{Haley2006}, design assumptions \cite{Roeller2006}, and construction assumptions \cite{Lehman2001}) have been extensively discussed. For instance, in the early phases of software development, there could be many uncertain things. However, in order to meet the project business goals (e.g., schedule and deadlines), stakeholders have to work in the presence of such uncertainties; these uncertainties can lead to assumptions. In this paper, we advocate treating uncertainty and assumption as two different but related concepts: one way to deal with uncertainties is to make implicit or explicit assumptions (i.e., a thing that is uncertain, but accepted as true)~\cite{Yangsms2018}. 

The importance of assumptions and their management in software development has been highlighted in many studies and industrial cases. For example, Corbat\'{o} \cite{Corbato1991} mentioned in his ACM Turing Award lecture that ``\textit{design bugs are often subtle and occur by evolution with early assumptions being forgotten as new features or uses are added to systems.}” Garlan et al. pointed out that incompatible assumptions in software architecture can cause architectural mismatch \cite{Garlan2009}. Lewis et al. also mentioned similar results in machine learning systems: since there are different types of stakeholders (e.g., data scientist, software engineer, and system user) of a machine learning system, they could make different but incompatible or invalid assumptions, leading to system misunderstanding, mismatch, etc \cite{lewis2021}.
In October 2018, Lion Air Flight 610 crashed 13 minutes after takeoff and killed all 189 people on board; In March 2019, Ethiopian Airlines Flight 302 crashed and ended another 157 lives. According to the reports from the government, one critical reason of the 737 MAX crashes is regarding not-well managed assumptions \cite{US2019}\cite{committee2020}. In the report, they mentioned that the aircraft company made invalid assumptions about the critical system components. Specifically, the invalid assumptions regarding MCAS (Maneuvering Characteristics Augmentation System) are the root cause of the crashes. The report also pointed out the need of identifying and re-evaluating important assumptions in the system. 

As evidenced from researchers and practitioners, stakeholders constantly make assumptions in their work \cite{Yangsms2018}\cite{Yangsurvey2016}. The assumptions can be further classified as two types: self-claimed assumptions (SCAs) and potential assumptions (PAs). 
Considering a sentence in a commit, pull request (PR), or issue of a GitHub repository, an SCA is that the sentence includes an assumption and the assumption is explicitly claimed using at least one of the assumption-related terms (i.e., ``assumption'', ``assumptions'', ``assume'', ``assumes'', ``assumed'', ``assuming'', ``assumable'', and ``assumably''). For example, in the sentence: ``\textit{[tf/xla] fixup numbering of xla parameters used for aliasing previously, the xla argument parameter was incorrectly assumed to be corresponding to the index in the vector of `xlacompiler::argument'}'', it includes an SCA of assuming the \textsc{xla argument} parameter is corresponding to the index in the vector of  \textsc{xlacompiler::argument} and the developer claimed that it was an invalid assumption. 
A PA is that the sentence may include an assumption, i.e., it is a potential assumption that needs further confirmation from human experts. This definition covers various aspects, such as expectations, future events, possibilities, guesses, opinions, feelings, and suspicions, which can indicate PAs. We provide three examples to further explain the PA concept. For example, in the sentence: ``\textit{i think the right way to create demo tensorboard instances is to simply run a tensorboard in the cloud, rather than keep maintaining this mocked-out backend.}'', it includes a PA regarding thoughts of the right way to create demo tensorboard instances. In another example: ``\textit{The system will not crash under heavy load}'', the sentence describes a future state of the system, which is uncertain, and includes a PA. The third example is: ``\textit{both false and true outputs should be considered independently}''. This sentence does not include assumption-related terms, but the sentence describes an expectation (i.e., ``something should be''), which is a PA. After further confirmation by human experts, this PA can be transformed to an SCA or other types of software artifacts.
Besides SCAs and PAs (which belong to explicit assumptions), there are also many implicit assumptions in projects (e.g., in stakeholders' heads or requiring reasoning). Identifying implicit assumptions at the sentence level is much more tricky than identifying SCAs and PAs, since there are no explicit clues in the sentences and stakeholders need to infer the sentences based on the context, which involves assumption reasoning. Identifying implicit assumptions is out of the scope of this study, and we treat it as future work.

Assumptions are related to many types of software artifacts, such as decisions, technical debt, and source code \cite{Yang2021}. For example, in TensorFlow, there is an SCA: ``\textit{TODO: Looks like there is an assumption that weight has only one user. We should add a check here}'', which induces a technical debt.
Existing research on assumptions and their management in software development usually use experiments, surveys, and case studies to manually identify and extract assumptions through observation, questionnaires, interviews, focus groups, and documentation analysis~\cite{Yangsms2018}. Since such approaches have high costs (e.g., time and resources), the number of identified and extracted assumptions in those studies is often limited, leading to various problems of developing new theories, approaches, and methods of assumptions and their management in software development.

In this work, to overcome the issues of manually identifying and extracting assumptions in software development, we proposed a tool: Assumption Miner, which can be used to automatically identify and extract assumptions (i.e., SCAs and PAs) on GitHub repositories. 
With over 100 million developers, 4 million organizations, and 330 million repositories\footnote{\url{https://github.com/about/}, accessed on 2023-04-21}, GitHub is one of the most important sources for open source software development. 
Besides assumptions, the tool can also be easily extended to other research fields (e.g., identifying and extracting technical debt~\cite{li2015systematic}). 

\textbf{How to access Assumption Miner}. Assumption Miner is available at \footnote{\url{http://39.108.224.140}}. Users can register or use a \textit{guest} account to login the tool. We also provided a deployment package for users who want to try the tool on their local environment \cite{deploy2023}. Users can read and follow the instructions in the description for the deployment of the tool.

The remainder of the paper is organized as follows. Section \ref{Related Work} provides related work, Section \ref{Assumption Miner} describes the details of Assumption Miner, Section \ref{Using Assumption Miner} presents an example of using the tool, Section \ref{Evaluation of Assumption Miner} describes an  evaluation of Assumption Miner, and Section \ref{Conclusions} concludes the paper with future directions.

\section{Related Work} \label{Related Work}
In the field of assumptions and their management in software development, most assumptions are manually identified and extracted by researchers and practitioners. Landuyt and Joosen focused on assumptions made during the application of a threat modeling framework (i.e., LINDDUN), which allows the identification of privacy-related design flaws in the architecting phase \cite{Landuyt2020}. They conducted a descriptive study with 122 master students, and the students identified and extracted 845 assumptions from the models created by the students. 
Yang et al. conducted an exploratory study of assumptions made in the development of nine popular deep learning frameworks (e.g., TensorFlow, Keras, and PyTorch) on GitHub \cite{Yang2021}. They identified and extracted 3,084 assumptions from the code comments in over 50,000 files of the deep learning frameworks.
Xiong et al. studied assumptions in the Hibernate developer mailing list, including their expression, classification, trend over time, and related software artifacts \cite{Xiong2018}. In their study, they identified and extracted 832 assumptions. 
Li et al. developed a machine learning approach \cite{Li2019} to identify and classify assumptions based on the dataset constructed by Xiong et al. \cite{Xiong2018}, which can read the data (i.e., sentences) from the dataset (i.e., a \textsc{.csv} file), preprocess the data (e.g., using NLTK and Word2Vec), train classifiers (e.g., Perception, Logistic Regression, and Support Vector Machines), and evaluate the trained classifiers (e.g., precision, recall, and F1-score). However, their approach is not specifically developed for PAs and SCAs and cannot mine assumptions from other sources (e.g., GitHub repositories).

Compared to the related work above, the tool (i.e., Assumption Miner) proposed in this work focuses on GitHub repositories, and can automatically collect data (e.g., issues, PRs, and commits) and identify and extract SCAs and PAs. Assumption Miner can also be easily extended to work with other repositories (e.g., Stack Overflow) or other types of software artifacts (e.g., technical debt).

\section{Assumption Miner} \label{Assumption Miner}
\begin{figure*} [h]
 \centering
  \includegraphics [scale=0.65] {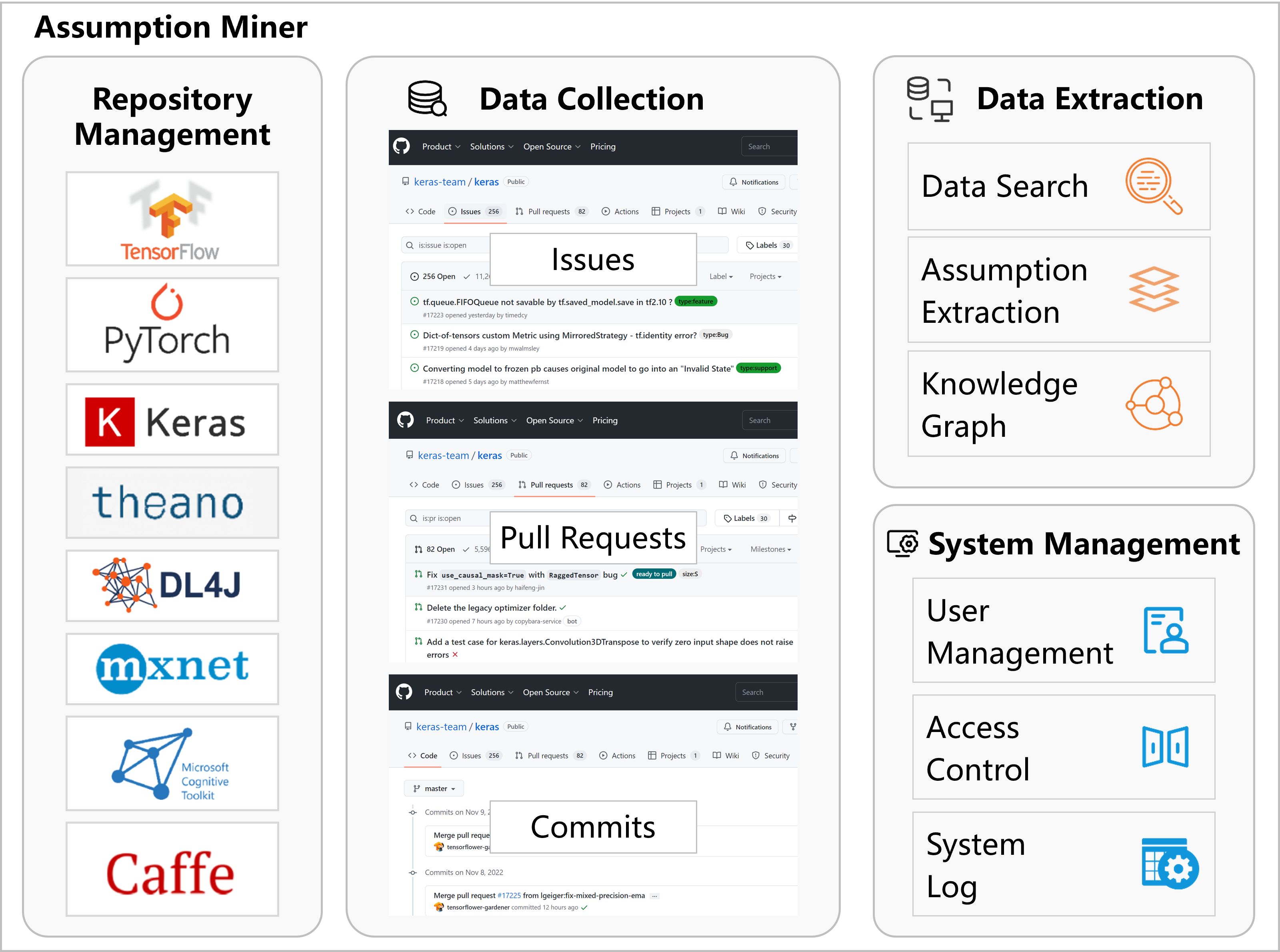}
 \caption{Modules of Assumption Miner}
 \label{modules}
\end{figure*}

Assumption Miner is composed of four modules: Repository Management, Data Collection, Data Extraction, and System Management, as shown in Fig. \ref{modules}.

\textbf{Repository Management} includes (1) getting information of the repositories and their releases from the GitHub server, (2) searching and showing details of the repositories, (3) downloading source code of each release of the repositories, and (4) deleting all the data of specific repositories. 
Adding a repository using Assumption Miner requires users to enter the owner (e.g., tensorflow) and name (e.g., tensorflow) of the repository. When adding a repository, Assumption Miner first checks whether the repository exists in the MySQL database and the GitHub server. If all the checks pass, Assumption Miner gets information (e.g., URL, releases, and tags) of the repository from the GitHub server and insert the data into the MySQL database. 

\textbf{Data Collection} aims to (1) show the data models of Repository, Release, Tag, PR, Commit, and Issue, (2) search and show data collection information of repositories and collect issues, PRs, and commits based on the data models, (3) monitor data collection processes, and (4) show data collection history. 
The data models are predefined and currently cannot be changed by users. The basic information of each repository (e.g., its releases and tags) and the information of the data collection processes are stored in a MySQL database, while the data of issues, PRs, and commits are stored in a MongoDB database.
When collecting data from the GitHub server using Assumption Miner, users can also set a time (default: 10 seconds) for automatically refreshing the data collection status (i.e., collecting, finished, and error). 
Assumption Miner uses cursors (i.e., issue cursor, PR cursor, and commit cursor) to record each batch of the data downloaded from the GitHub server, and therefore Assumption Miner supports continuing data collection after it has been stopped due to errors and exceptions (e.g., over the limits by GitHub).

\textbf{Data Identification and Extraction} is composed of three submodules: Assumption Extraction, Data Search, and Knowledge Graph. 
\textbf{Assumption Extraction} contains four functions: SCA Identification, SCA Extraction, PA Identification, and PA Extraction. 
In the \textbf{\textit{SCA Identification}} function, we used a keyword-based search approach for identifying SCAs (i.e., word level), based on the \textit{assumption} related search terms (i.e., \textit{assumption}, \textit{assumptions}, \textit{assume}, \textit{assumes}, \textit{assumed}, \textit{assuming}, \textit{assumable}, and \textit{assumably}) and the following search scope: (1) title, body, body of comments of issues, (2) title, body, body of comments of PRs, and (3) message of commits. 
Since the results from using the SCA Identification function are at the word level (i.e., highlighting the matched terms), in the \textbf{\textit{SCA Extraction}} function, Assumption Miner provides support for locating, matching, and extracting the related sentences that include the matched terms (i.e., at the sentence level). As most of the description (e.g., description of an issue) is created and edited by GitHub users, there could be punctuation problems existing in the description (e.g., a sentence does not have a ``." or ``." is used in the source code, such as ``a.b"), which may lead to errors in the separation of the sentences. Therefore, we manually identified the patterns of SCA description from the projects on GitHub, and implemented the patterns in the SCA Extraction function to extract SCA sentences. 
The outputs of SCA Extraction contain a set of data items as shown in Table \ref{Data Items of Issues and PRs in SCA Extraction} and Table \ref{Data Items of Commits in SCA Extraction}.
Moreover, there are sentences without using \textit{assumption} related keywords, but could act as assumptions (we call them potential assumptions, PAs). Though PAs are not SCAs, they can be further reviewed by stakeholders and transformed to SCAs (as the inputs for SCA extraction). Therefore, Assumption Miner also provides support for identifying and extracting such assumptions through the \textbf{\textit{PA Identification}} and \textbf{\textit{PA Extraction}} function, which complements SCA identification and extraction. The first author manually collected and labeled 35,855 sentences from the issues, PRs, and commits of multiple repositories (e.g., Keras and Theano), following the guidelines of assumption identification proposed in \cite{Yangsms2018}. Then the other authors reviewed the results and reached a consensus with the first author. After assumption collection and labeling, we constructed a dataset for PAs, fine-tuned a deep learning model based on ALBERT (a lite BERT, which architecture is based on BERT), and trained and adjusted a classification model for PA identification \cite{Lan2020}. The reason of choosing ALBERT is because ALBERT is one of the most powerful language models, which can achieve good performance with fewer parameters (compared to BERT) in many tasks, such as the binary single-sentence classification task \cite{Lan2020}. 
Since we are using deep learning models and identifying PAs at the sentence level, the identification process could be rather slow on CPUs (e.g., it may take hours/days on our server to identify PAs, depending on the amount of data to be processed). Therefore, we used a queue (i.e., a waiting list with a first-in first-out strategy) on the server to manage the tasks of identifying PAs. In the PA Extraction function, Assumption Miner organizes the PAs (sentences) identified from the PA Identification function into a file (similar to SCA Extraction), and users can download the file for further review. 

\begin{table}[h]
\centering
\caption{Data Items of Issues and PRs in SCA Extraction}
\label{Data Items of Issues and PRs in SCA Extraction}
\begin{tabular}{|c|c|}
\hline
\textbf{Data Item} & \textbf{Description} \\ \hline
{owner} & {The owner of the repository} \\ \hline
{repo\_name} & {The name of the repository} \\ \hline
{author} & {The author of the issue or PR} \\ \hline
{title} & {The title of the issue or PR} \\ \hline
{state} & {The state of the issue or PR} \\ \hline
{url} & {The URL of the issue or PR} \\ \hline
{SCA} & {The SCA sentence in the issue or PR} \\ \hline
\end{tabular}
\end{table}

\begin{table}[h]
\centering
\caption{Data Items of Commits in SCA Extraction}
\label{Data Items of Commits in SCA Extraction}
\begin{tabular}{|c|c|}
\hline
\textbf{Data Item} & \textbf{Description} \\ \hline
{owner} & {The owner of the repository} \\ \hline
{repo\_name} & {The name of the repository} \\ \hline
{author\_name} & {The author of the issue or PR} \\ \hline
{message} & {The message of the issue or PR} \\ \hline
{url} & {The URL of the issue or PR} \\ \hline
{SCA} & {The SCA sentence in the issue or PR} \\ \hline
\end{tabular}
\end{table}

\textbf{Data Search} aims to search and show specific issues, PRs, and commits based on keywords. 
In data search, Assumption Miner requires users to specify which repository (e.g., \textit{TensorFlow}), data type (i.e., \textit{issue, PR}, and \textit{commit}), search scope (e.g., \textit{title}), and keywords (e.g., \textit{assume}) to search. For example, for issues of the TensorFlow project, a search scope can be \textit{title body comments.body}, which means that Assumption Miner will search data within the scope of the title of the issues, the body of the issues, and the body of the comments of the issues in the TensorFlow project. 
For keywords, Assumption Miner supports \textsc{AND} (i.e., using double quotation marks, e.g., \textit{``assume" ``software"}) and \textsc{OR} (i.e., without quotation marks, e.g., \textit{assume software}).
The search terms are highlighted in the search results. If the description of a data item is too long, users can click on the ``detail" button to see the full information of the item.

\textbf{Knowledge Graph} supports both traditional knowledge graph and dynamic knowledge graph. Assumption Miner provides three dimensions (i.e., release, month, and day) to construct the timeline of the data. For each repository, Assumption Miner creates and connects entities according to the timeline and their states. For example, a PR can be published, merged, and closed, and therefore Assumption Miner creates three connected entities if they are within a timeline. 

\textbf{System Management} supports user registration, login, and logoff, and provides access control and system logs.

\begin{figure*} [h]
 \centering
  \includegraphics [scale=0.24] {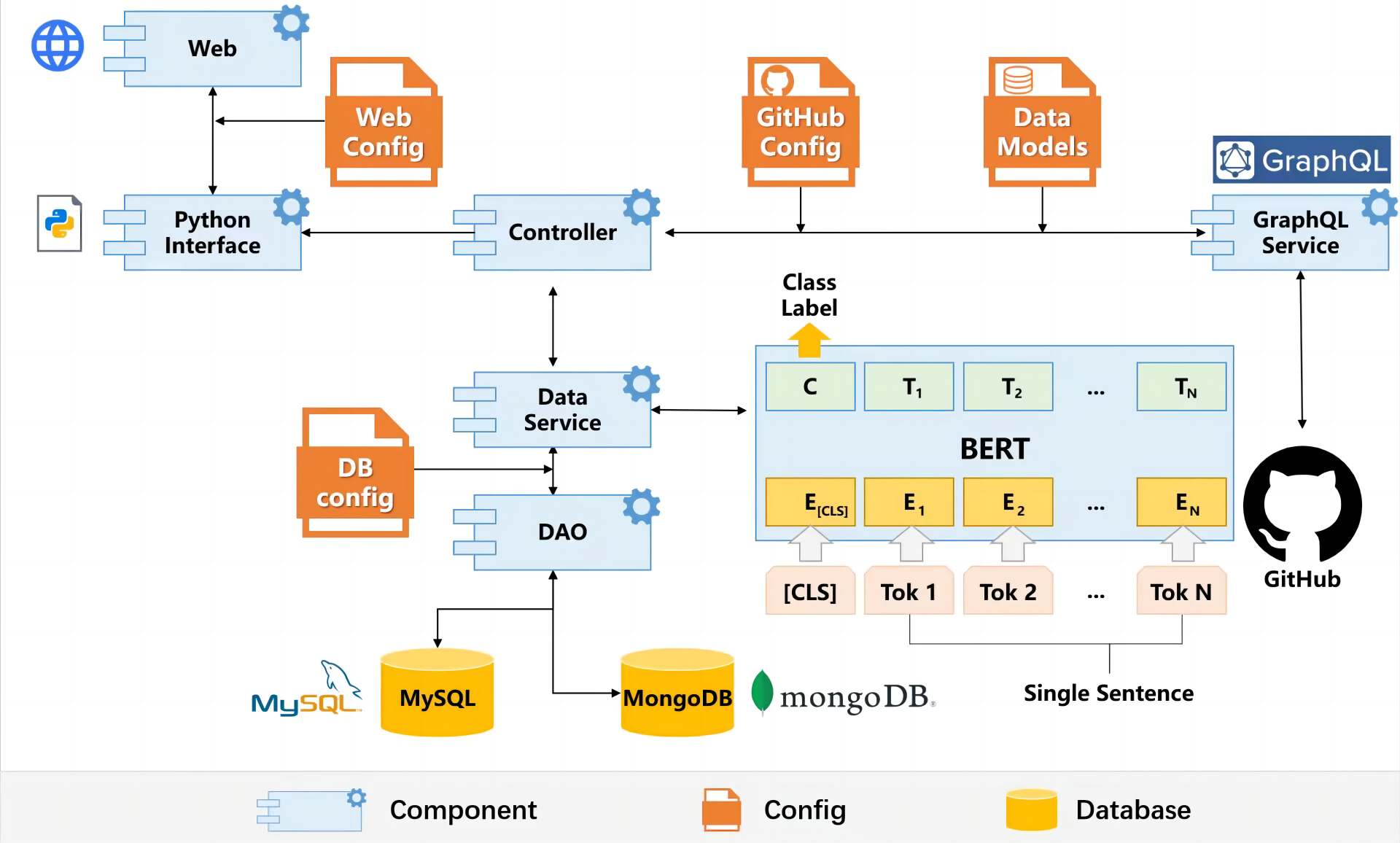}
 \caption{Architecture of Assumption Miner}
 \label{architecture}
\end{figure*}

The architecture of Assumption Miner is shown in Fig. \ref{architecture}. When a user clicks on a menu or button, the Web component organizes the data, generates a request, and sends it to the Controller component through the Python Interface component. The Controller component analyzes the request: 
(1) If the request is regarding getting data from the GitHub server, the Controller component organizes the data and calls the functions in the GitHub Service component. The GitHub Service component reads the GitHub configuration (stored in the system) and organizes queries based on the predefined data models. Then the GitHub Service component sends requests to the GitHub server, gets responses from the GitHub server, analyzes the responses, and sends back the data to the Controller component. 
(2) If the request is regarding interacting with the MySQL or the MongoDB database, the Controller component organizes the data and calls the functions in the Data Service component. The Data Service component further organizes the data and calls the functions in the DAO (Data Access Object) component, which implements the interaction with the MySQL or the MongoDB database. The DAO component reads the database configuration (stored in the system), communicates with the databases, gets the results from the databases, and sends them back to the Data Service component. The Data Service component sends the data getting from the databases back to the Controller component. 
(3) If the request is regarding using the trained model (based on ALBERT) to identify PAs, the Controller component organizes the data and calls the functions in the Data Service component. The Data Service component preprocesses the data, loads the trained model, and uses the model to identify PAs. The results are then sent back to the Controller component. 
Finally, the Controller component organizes the data getting from GitHub, the MySQL database, the MongoDB database or the deep learning model, sends the data back to the Web component through the Python Interface component, and then the Web component shows the results to the user.

\section{Using Assumption Miner} \label{Using Assumption Miner}
In this section, we walk the usage of Assumption Miner through an example: the TensorFlow project on GitHub \footnote{\url{https://github.com/tensorflow/tensorflow}}. TensorFlow is one of the most popular deep learning frameworks, which is widely used in many deep learning systems and application domains. 
The TensorFlow project on GitHub started in 2015, having 188 releases \footnote{\url{https://github.com/tensorflow/tensorflow/releases} (accessed on 2023-04-21)}, 146,893 commits\footnote{\url{https://github.com/tensorflow/tensorflow} (accessed on 2023-04-21)}, 22,887 PRs \footnote{\url{https://github.com/tensorflow/tensorflow/pulls?q=is\%3Apr} (accessed on 2023-04-21)}, and 37,074 issues \footnote{\url{https://github.com/tensorflow/tensorflow/issues?q=is\%3Aissue} (accessed on 2023-04-21)} till April 2023.
Users need to register an account and set a personal access token of GitHub \footnote{\url{http://www.m58.link/cxaNX}}
when using Assumption Miner. The token is used to access the GitHub Application Programming Interface (API), since Assumption Miner needs to communicate with the GitHub API to get data (e.g., issues, PRs, and commits). We also provide a default token for Assumption Miner users. However, since GitHub has limitations in place to protect against excessive or abusive calls to GitHub servers (e.g., the rate limit is 5,000 points per hour and individual calls cannot request more than 500,000 total nodes)\footnote{\url{https://docs.github.com/en/graphql/overview/resource-limitations}}, using the default token may lead to errors in data collection because of these limitations. 
After registration of the Assumption Miner account, users can login Assumption Miner with the account. 
Below is the process of using Assumption Miner to identify and extract assumptions from the TensorFlow project on GitHub.

\textbf{Create the TensorFlow repository}.
Users need to click on the Repository Management module, then click on the ``Add" button, enter the owner as ``tensorflow" and the name as ``tensorflow", and click on the ``Save" button to create the TensorFlow repository on Assumption Miner. For each release of a repository, Assumption Miner provides users a link to download the source code in the Repository Management module (this is an optional step). Then users can use tools such as Visual Studio Code and PyCharm to further browse the code and search assumptions in the code. 

\textbf{Collect issues, PRs, and commits on TensorFlow}.
After the TensorFlow repository is created on Assumption Miner, users can further use the Data Collection module to collect issues, PRs, and commits of the TensorFlow repository. 
Users can start multiple tasks simultaneously, but this could cause errors because of the limitation by GitHub.

\textbf{Identify and Extract SCAs on TensorFlow}.
In the Assumption Extraction submodule of the Data Extraction module, users need to select the \textit{TensorFlow} repository and a data type, and click on the ``SCA Identification" button. Assumption Miner will show the results and highlight all the SCAs. 
Users can further extract the data to a \textsc{csv} file to construct a dataset by clicking on the ``SCA Extraction" button. The first line in the \textsc{csv} file is the title, indicating the repository and type (i.e., issue, PR, and commit) of the extracted data.
If users want to search data on the TensorFlow repository (optional step), they can use the Data Search submodule of the Data Extraction module, by selecting the repository (e.g., ``TensorFlow") and the data type (e.g., ``issue") and specifying the search scope (e.g., ``title") and the search term (e.g., ``assumption"). 

\textbf{Identify and Extract PAs on TensorFlow}.
In the Assumption Extraction submodule of the Data Extraction module, users need to select the \textit{TensorFlow} repository and a data type, and click on the ``PA Identification" button. Assumption Miner will show the results and highlight all the sentences that may include a PA. After PA identification, users can click on the ``PA Extraction" button to download the results of PA identification for further review.

\textbf{Generate a knowledge graph of the assumptions on TensorFlow}.
This step is optional. Users need to select the \textit{TensorFlow} repository and a dimension (i.e., \textit{release}, \textit{month}, or \textit{day}) to construct a knowledge graph of assumptions based on the chosen dimension. 

The aforementioned results can be further used in various context. For example, through identifying assumptions, users may better understand what was assuming in a certain project and deal with such uncertainty in their future work.

\section{Evaluation of Assumption Miner} \label{Evaluation of Assumption Miner}
\begin{figure*} [h]
 \centering
  \includegraphics [scale=0.15] {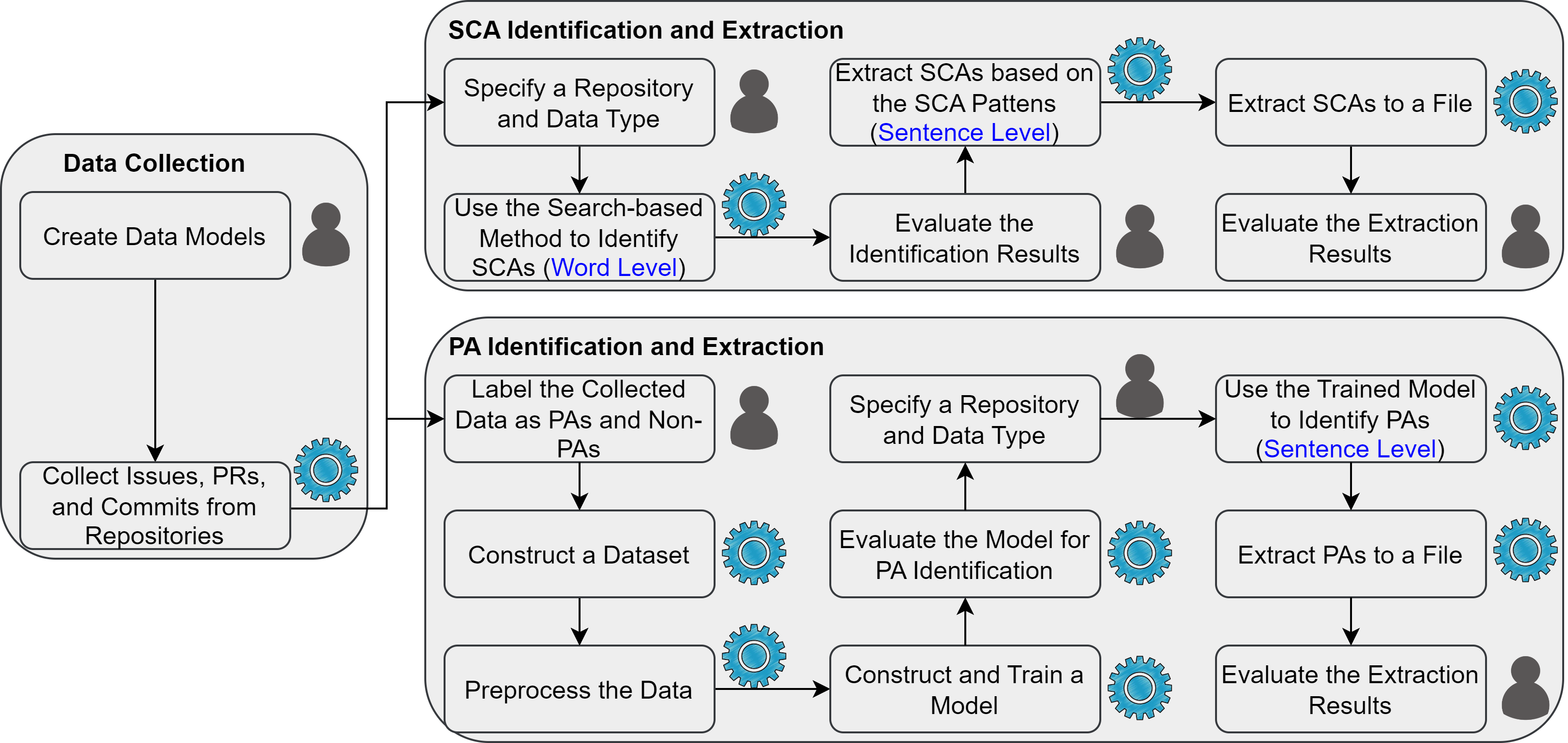}
 \caption{Evaluation of Assumption Miner}
 \label{evaluation}
\end{figure*}

We conducted an evaluation on data collection, SCA identification, SCA extraction, PA identification, and PA extraction, as shown in Fig. \ref{evaluation}. 
The output of data collection is the input of SCA identification and PA identification, the output of SCA identification is the input of SCA extraction, and the output of PA identification is the input of PA extraction.

\subsection{Evaluation of Data Collection} \label{Evaluation of Data Collection}
We conducted an evaluation of using Assumption Miner to collect issues, PRs, and commits on seven GitHub repositories: Caffe, CNTK, Theano, DeepLearning4J (DL4J), MXNet, Keras, and TensorFlow, regarding the completeness and performance of Assumption Miner on data collection. We first created an issue data model (including \textit{repository name}, \textit{title}, \textit{ID}, \textit{author}, \textit{URL}, \textit{labels}, \textit{state}, \textit{body}, and \textit{comments}), a PR data model (including \textit{repository name}, \textit{owner}, \textit{title}, \textit{ID}, \textit{author}, \textit{URL}, \textit{labels}, \textit{state}, \textit{body}, \textit{comments}, and \textit{reviews}), and a commit data model (including \textit{repository name}, \textit{owner}, \textit{OID}, \textit{author name}, \textit{author email}, \textit{committed date}, \textit{URL}, and \textit{message}). The data items of each data model were selected from the GraphQL API on GitHub\footnote{\url{https://docs.github.com/en/graphql/reference/objects}}. 

The configuration of the server we used for the evaluation of the data collection is: (1) CPU: Intel(R) Xeon(R) Platinum 8255C CPU @ 2.50GHz, 2 cores, (2) Memory: 2GB, (3) 
Hard Disk: 40GB SSD, (4) Operation System: Linux VM-20-4-centos 3.10.0-1160.45.1.el7.x86\_64 \#1 SMP. The results of using Assumption Miner to collect data are shown in Table \ref{Completeness and Performance of Data Collection of using Assumption Miner}.

\begin{table*}[h]
\centering
\caption{Completeness and Performance of Data Collection of using Assumption Miner}
\label{Completeness and Performance of Data Collection of using Assumption Miner}
\begin{tabular}{|c|c|c|c|c|c|}
\hline
\textbf{Repository} & \textbf{Data Items on GitHub} & \textbf{Collected Data Items} & \textbf{Data Type} & \textbf{Time (s)} & \textbf{Access Date} \\ \hline
{Caffe} & {4,786} & {4,786} & {Issue} & {400} & {2022/11/4} \\ \hline
{Caffe} & {2,238} & {2,238} & {PR} & {152} & {2022/11/4} \\ \hline
{Caffe} & {4,156} & {4,156} & {Commit} & {51} & {2022/11/4} \\ \hline
{CNTK} & {3,288} & {3,288} & {Issue} & {228} & {2022/11/4} \\ \hline
{CNTK} & {557} & {557} & {PR} & {27} & {2022/11/4} \\ \hline
{CNTK} & {16,117} & {16,117} & {Commit} & {181} & {2022/11/4} \\ \hline
{Theano} & {2,671} & {2,671} & {Issue} & {246} & {2022/11/4} \\ \hline
{Theano} & {4,114} & {4,114} & {PR} & {198} & {2022/11/4} \\ \hline
{Theano} & {28,127} & {28,127} & {Commit} & {335} & {2022/11/4} \\ \hline
{DL4J} & {5,652} & {5,652} & {Issue} & {290} & {2022/11/7} \\ \hline
{DL4J} & {4,185} & {4,185} & {PR} & {140} & {2022/11/7} \\ \hline
{DL4J} & {2,606} & {2,606} & {Commit} & {25} & {2022/11/7} \\ \hline
{MXNet} & {9,532} & {9,532} & {Issue} & {465} & {2022/11/8} \\ \hline
{MXNet} & {11,090} & {11,090} & {PR} & {551} & {2022/11/8} \\ \hline
{MXNet} & {11,893} & {11,893} & {Commit} & {145} & {2022/11/7} \\ \hline
{Keras} & {11,518} & {11,518} & {Issue} & {519} & {2022/11/8} \\ \hline
{Keras} & {5,670} & {5,670} & {PR} & {208} & {2022/11/8} \\ \hline
{Keras} & {7,493} & {7,493} & {Commit} & {65} & {2022/11/8} \\ \hline
{TensorFlow} & {35,966} & {35,966} & {Issue} & {2,402} & {2022/11/9} \\ \hline
{TensorFlow} & {22,119} & {22,119} & {PR} & {1,105} & {2022/11/9} \\ \hline
{TensorFlow} & {138,366} & {138,366} & {Commit} & {1,295} & {2022/11/9} \\ \hline
\end{tabular}
\end{table*}

The results show that Assumption Miner can effectively collect issues, PRs, and commits of the repositories on GitHub. Certain repositories (e.g., TensorFlow) may be frequently updated, and there is a need to construct a mechanism for continuously collecting data.

\subsection{Evaluation of SCA Identification} \label{Evaluation of SCA Identification}
After data collection (as mentioned in Section \ref{Evaluation of Data Collection}), we used the collected data (i.e., issues, PRs, and commits) of the Keras and TensorFlow repository to conduct an evaluation on identifying SCAs using Assumption Miner (i.e., the SCA Identification function).   
The first author manually checked the identified results to classify them as SCAs or non-SCAs. The evaluation results of SCA identification are shown in Table \ref{Results of Identifying SCAs}. 

The count of the identified SCAs could be larger than the search results (e.g., count of messages in the commits of the Keras repository), since each issue, PR, or commit may include multiple SCAs. For example, an issue \footnote{\url{https://github.com/keras-team/keras/issues/395}} 
of Keras mentions: ``\textit{Assume we are trying to learn a sequence to sequence map. For this we can use Recurrent and TimeDistributedDense layers. Now assume that the sequences have different lengths. We should pad both input and desired sequences with zeros, right? But how will the objective function handle the padded values? There is no choice to pass a mask to the objective function. Won't this bias the cost function?}", which includes two SCAs: ``assume we are trying to learn a sequence to sequence map" and ``assume that the sequences have different lengths".

Since Assumption Miner used a keyword-based (i.e., the \textit{assumption} related terms) search approach for SCA identification, it could go wrong in certain context (e.g., a variable in a code snippet named \textit{assume}). 
Moreover, we also found that certain SCAs lack details. For example, an issue \footnote{\url{https://github.com/keras-team/keras/issues/1174}} 
of Keras mentioned: ``\textit{strict enforcement of user input assumptions, and raising of helpful error messages.}" However, we cannot understand what exactly the user input assumptions are. These SCAs need to be further processed by Assumption Miner (e.g., add warnings in the results).

\begin{table*}[h]
\centering
\caption{Results of Identifying SCAs using Assumption Miner}
\label{Results of Identifying SCAs}
\begin{tabular}{|c|c|c|c|c|c|}
\hline
\textbf{Repository} & \textbf{Data Type} & \textbf{Search Field} & \textbf{Search Results} & \textbf{Identified SCAs} & \textbf{Misidentification} \\ \hline
{Keras} & {Issue} & {title} & {3} & {3} & {0} \\ \hline
{Keras} & {Issue} & {body} & {253} & {298} & {13} \\ \hline
{Keras} & {PR} & {title} & {3} & {3} & {0} \\ \hline
{Keras} & {PR} & {body} & {59} & {65} & {0} \\ \hline
{Keras} & {Commit} & {message} & {10} & {11} & {0} \\ \hline
{TensorFlow} & {Issue} & {title} & {13} & {13} & {0} \\ \hline
{TensorFlow} & {Issue} & {body} & {662} & {797} & {47} \\ \hline
{TensorFlow} & {PR} & {title} & {5} & {5} & {0} \\ \hline
{TensorFlow} & {PR} & {body} & {136} & {150} & {6} \\ \hline
{TensorFlow} & {Commit} & {message} & {567} & {616} & {39} \\ \hline
\multicolumn{3}{|r|}{\textbf{Total}} & {1,711} & {1,961} & {105} \\ \hline
\end{tabular}
\end{table*}

Regarding the performance of the SCA Identification process, we used an Aliyun server \footnote{\url{https://www.alibabacloud.com/en}} with the following configuration: (1) Server type: ecs.s6-c1m2.xlarge, (2) CPU: 4 cores (vCPU), (3) Memory: 8GB, (4) Hard Disk: 40GB SSD, (5) Operation System: Windows Server 2022,  and conducted an evaluation on the issues, PRs, and commits of Keras and Tensorflow. The results are shown in Table \ref{sca identification performance}.

\begin{table}[h]
\centering
\caption{Performance of the SCA Identification Process}
\label{sca identification performance}
\begin{tabular}{|c|c|c|c|}
\hline
\textbf{Repository} & \textbf{Data Type} & \textbf{Identification Field} & \textbf{Time (s)} \\ \hline
{Keras} & {Issue} & {title, body, comments.body} & {7.287} \\ \hline
{Keras} & {PR} & {title, body, comments.body} & {0.650} \\ \hline
{Keras} & {Commit} & {message} & {0.378} \\ \hline
{TensorFlow} & {Issue} & {title, body, comments.body} & {7.419} \\ \hline
{TensorFlow} & {PR} & {title, body, comments.body} & {0.632} \\ \hline
{TensorFlow} & {Commit} & {message} & {0.681} \\ \hline
\end{tabular}
\end{table}

The results show that Assumption Miner can correctly identify 94.92\% SCAs (1,961 SCAs out of 2,066 SCAs) from the issues, PRs, and commits of the Keras and TensorFlow repository. Certain variables and functions named as for example \textit{assume} exist in issues, PRs, and commits, which needs to be further processed by Assumption Miner.

\subsection{Evaluation of SCA Extraction}
We further evaluated whether Assumption Miner (i.e., the SCA Extraction function) can correctly extract SCAs (i.e., at the sentence level) using the identification results (i.e., 1961 identified SCAs) of the Keras and TensorFlow repository from \ref{Evaluation of SCA Identification}. The first author manually checked the extracted results to classify them as correct extraction and missed extraction. The results are shown in Table \ref{Results of Extracting SCAs}. As an example, there are 298 SCAs in the body of Keras issues. Assumption Miner correctly extracted 290 of the SCAs, but missed 8 SCAs.

\begin{table}[h]
\centering
\caption{Results of Extracting SCAs using Assumption Miner}
\label{Results of Extracting SCAs}
\begin{tabular}{|p{0.15\columnwidth}|p{0.09\columnwidth}|p{0.12\columnwidth}|p{0.12\columnwidth}|p{0.11\columnwidth}|p{0.10\columnwidth}|}
\hline
\textbf{Repository} & \textbf{Data Type} & \textbf{Extraction Field} & \textbf{SCAs} & \textbf{Correct} & \textbf{Missed} \\ \hline
{Keras} & {Issue} & {title} & {3} & {3} & {0} \\ \hline
{Keras} & {Issue} & {body} & {298} & {290} & {8} \\ \hline
{Keras} & {PR} & {title} & {3} & {3} & {0} \\ \hline
{Keras} & {PR} & {body} & {65} & {65} & {0} \\ \hline
{Keras} & {Commit} & {message} & {11} & {11} & {0} \\ \hline
{TensorFlow} & {Issue} & {title} & {13} & {13} & {0} \\ \hline
{TensorFlow} & {Issue} & {body} & {801} & {772} & {29} \\ \hline
{TensorFlow} & {PR} & {title} & {5} & {5} & {0} \\ \hline
{TensorFlow} & {PR} & {body} & {150} & {146} & {4} \\ \hline
{TensorFlow} & {Commit} & {message} & {616} & {609} & {7} \\ \hline
\multicolumn{3}{|r|}{\textbf{Total}} & {1,961} & {1,913} & {48} \\ \hline
\end{tabular}
\end{table}

The results show that Assumption Miner can correctly extract 97.55\% SCAs (1,913 SCAs out of 1,961 SCAs) from the issues, PRs, and commits of the Keras and TensorFlow repository. Certain structures of the issues, PRs, and commits (e.g., ``assume" and ``assumption" exist in one sentence) may lead to errors in SCA extraction, which needs further investigation and improvements.

\subsection{Evaluation of PA Identification}
For PAs, we manually labeled 35,855 sentences from the issues, PRs, and commits of multiple repositories (e.g., Keras and Theano), and constructed a dataset, containing a training set and a test set with a data proportion of 8:2 from the labeled sentences. We created a vocabulary and tokenized the data from the dataset based on the vocabulary. Then we constructed the deep learning model (based on ALBERT), trained the model for 50,000 epochs with a batch size of 32 and a learning rate of 2e-5. 

Regarding the performance of the PA identification process, we used the same configuration used in the evaluation of SCA identification (see Section \ref{Evaluation of SCA Identification}), and conducted an evaluation on the issues, PRs, and commits of Keras and Tensorflow. The results are shown in Table \ref{pa identification performance}.

\begin{table}[h]
\centering
\caption{Performance of the PA Identification Process}
\label{pa identification performance}
\begin{tabular}{|c|c|c|c|}
\hline
\textbf{Repository} & \textbf{Data Type} & \textbf{Identification Field} & \textbf{Time} \\ \hline
{Keras} & {Issue} & {title, body, comments.body} & {11h 12m 58s} \\ \hline
{Keras} & {PR} & {title, body, comments.body} & {2h 28m 6s} \\ \hline
{Keras} & {Commit} & {message} & {24m 4s} \\ \hline
{TensorFlow} & {Issue} & {title, body, comments.body} & {47h 2m 34s} \\ \hline
{TensorFlow} & {PR} & {title, body, comments.body} & {12h 16m 59s} \\ \hline
{TensorFlow} & {Commit} & {message} & {10h 19m 44s} \\ \hline
\end{tabular}
\end{table}

Using GPU would significantly improve the performance of the PA identification process. As an example, we used NVIDIA GeForce RTX 3060 Ti to run the PA identification process on Keras. The results are 16 minutes 52 seconds (compared to 11 hours 12 minutes 58 seconds using CPU) on issues, 3 minutes 51 seconds (compared to 2 hours 28 minutes 6 seconds using CPU) on PRs, and 43 seconds (compared to 24 minutes 4 seconds using CPU) on commits of Keras.

The results show that the best accuracy of identifying PAs using Assumption Miner on the test set is 0.9451 on Epoch 22,000. Considering repositories may have various context (e.g., different development policies), there is a need to further extend the dataset to include more data from the repositories, which will help to improve the generalization ability of Assumption Miner in identifying PAs.

\section{Conclusions} \label{Conclusions}
Assumptions and their management are important in software development. The prerequisite of analyzing and understanding assumptions in software development is to identify and extract those assumptions with acceptable effort. To this end, we proposed Assumption Miner to automatically identify and extract assumptions on GitHub projects. Besides providing a running example of using Assumption Miner on the TensorFlow project, we also evaluated the performance of Assumption Miner, and the results show that Assumption Miner can effectively identify and extract assumptions from the repositories on GitHub. 
Assumption Miner can be potentially used for the research topics regarding assumptions and their management in software development, such as assumption making, evolution, evaluation, and reasoning.

For future work, the following aspects of Assumption Miner can be further optimized: (1) There is a need to construct a mechanism for continuously collecting data using Assumption Miner. (2) The identification of SCAs and PAs can be further optimized (e.g., develop new deep learning models, construct a larger dataset and train deep learning models based on the dataset). (3) Certain patterns of the issues, PRs, and commits (e.g., a variable named ``assume") may lead to incorrect SCA identification and extraction, which can be further addressed in Assumption Miner. (4) Assumptions are related to various types of software artifacts (e.g., requirements, design decisions, and technical debt). Automatically recovering the relationships between assumptions and such artifacts in Assumption Miner is a promising future direction. (5) Besides SCAs and PAs, there are many implicit assumptions in projects, which should also be identified and extracted in the future.

\section*{Acknowledgments}
This work is funded by Shenzhen Polytechnic with Grant No. 6022312043K, State Key Laboratory for Novel Software Technology at Nanjing University with Grant No. KFKT2022B37, and the National Natural Science Foundation of China (NSFC) with Grant No. 62172311.

\balance

\bibliographystyle{IEEEtran}
\bibliography{ref}

\end{document}